\documentclass[12pt, a4paper, oneside]{article}
\usepackage[latin1]{inputenc}
\usepackage{multicol}
\usepackage{color}
\usepackage{graphicx}
\usepackage[top=4cm,bottom=4cm,left=2.5cm,right=2.5cm]{geometry}
\usepackage[T1]{fontenc}
\usepackage[affil-it]{authblk}
\usepackage{cite}


\newcommand*\samethanks[1][\value{footnote}]{\footnotemark[#1]}

\newcommand{\beginsupplement}{%
        \setcounter{figure}{0}
        \renewcommand{\thefigure}{S\arabic{figure}}%
     }

\begin{document}
\title{\large \bf Curvature-driven, One-step Assembly of Reconfigurable Smectic Liquid Crystal  ``Compound Eye'' Lenses}
\date{}
\author[1]{Francesca Serra\thanks{These authors contributed equally to the work}}
\author[1]{Mohamed A. Gharbi \samethanks}
\author[2]{Yimin Luo}
\author[2]{Iris B. Liu}
\author[2]{Nathan D. Bade}
\author[1]{Randall D. Kamien}
\author[3]{Shu Yang}
\author[2]{Kathleen J. Stebe\thanks{kstebe@seas.upenn.edu}}
\affil[1]{Dept. Physics and Astronomy, University of Pennsylvania}
\affil[2]{Dept. Chemical and Biomolecular Engineering, University of Pennsylvania}
\affil[3]{Dept. Materials Science and Engineering, University of Pennsylvania}

\maketitle

{Confined smectic A liquid crystals (SmA LCs) form topological defects called focal conic domains (FCDs)\cite{DeGennes} that focus light as gradient-index lenses\cite{kimmicrolens1, kimmicrolens2, kimphotomask, kimreview}. Here, we exploit surface curvature to self-assemble FCDs in a single step into a hierarchical structure \cite{beller, meyer} (coined  ``flower pattern'') molded by the fluid interface that is pinned at the top of a micropillar. 
The structure resembles the compound eyes of some invertebrates, which consist of hundreds of microlenses on a curved interface, able to focus and construct images in three dimensions (3D) \cite{land}.  Here we demonstrate that these flowers are indeed  ``compound eyes'' with important features which  have not been demonstrated previously in the literature.  The eccentric FCDs gradually change in size with radial distance from the edge of the micropillar, resulting in a variable microlens focal length that ranges from a few microns to a few tens of microns within a single  ``flower''. We show that the microlenses can construct a composite 3D image from different depth of field (DOF). Moreover, the smectic ``'compound eye'' can be reconfigured by heating and cooling at the LC phase transition temperature; its field of view (FOV) can be manipulated by tuning the curvature of the LC interface, and the lenses are sensitive to light polarization. \linebreak[4]}

Insects' eyes are comprised of hundreds of microlenses (ommatidia) arranged on a curved surface \cite{land}. Despite having modest resolution in comparison to single aperture lenses (like the human eyes), the compound eye offers attractive optical properties, including exceptionally wide FOV, fast motion detection, and polarization sensitivity. Artifical compound eyes have been created with angular sensitivity\cite{lee} and with a hemispheric FOV and near infinite DOF \cite{rogers}. Typically, multiple  top-down fabrication steps are required, including photolithography, replica molding, or complex micromachining processes\cite{lee, rogers, dong}. For practical applications with wide FOV, lenses which self-align are highly desirable. Furthermore,  all artificial compound eyes demonstrated in the literature lack polarization sensitivity, which is an important attribute of compound eyes vs. human eyes, allowing for insects tracking and navigation \cite{fosterbio, homberg}. Finally, given the continued miniaturization  of optical devices, and that the typical size of ommatidia is tens of microns, it will be highly attractive to be able to self-assemble microlens arrays with wide and tunable FOV and polarization sensitivity. 
Among many soft materials, liquid crystals (LCs) are known for their optical anisotropy, sensitivity to light polarization, and reconfigurability near the phase transition temperature. Aside from the display industry, there has been a growing interest in using liquid crystals to make more sophisticated optical objects like sensors \cite{abbott}, lasers \cite{finkelmann}, and microlenses \cite{masuda}. 
Recently, we and others have used SmA LCs as topographical templates \cite{honglawan1, honglawan2, smalyukh}, photomasks,\cite{kimphotomask} and lenses\cite{kimmicrolens1, kimmicrolens2}. SmA LCs spontaneously form geometric defects called focal conic domains (FCDs) that act as microlenses and self-assemble into a wide range of arrays\cite{kimreview}. Within each FCD the SmA LC molecular layers bend to form the singular points, which are all arranged in two conics: an ellipse at the basis of the FCD, and a branch of hyperbola that passes through one focus of the ellipse\cite{DeGennes}. FCDs are actually gradient refractive index microlenses whose size and shape can be controlled by the thickness of smectic layers \cite{kimmicrolens1, kimmicrolens2, kimphotomask, kimreview}, surface anchoring and surface topography.  \linebreak[4]

When a thin smectic film with hybrid boundary conditions (i.e., the director is parallel to one substrate and perpendicular to the other) is pinned at the surface of a colloid, FCDs are generated, creating a  ``flower pattern'' \cite{beller}, in which FCDs are arranged like flower petals surrounding the colloid: large FCDs are observed close to the colloid, while smaller ``petals'' appear in the low curvature region away from the colloid. Spatially varying FCD sizes arise as a result of varying curvatures. The eccentricity of FCDs, defined as the displacement of the defect core (the intersection of the branch of hyperbola with the ellipse) from the geometrical center of the FCD (the center of the ellipse), also varies with distance from the colloid. The structure of the SmA LC layers in the flower pattern has been characterized and the mechanism of formation of the flower patten has been proposed by Beller et al.\cite{beller}. Yet, it is not clear whether such smectic flower would focus light and construct 3D images. To better control the size and eccentricity of FCDs and, more importantly, to exploit the unique optical properties of the SmA LCs, here, we cast a thin layer of SmA LC film around a transparent micropillar array, which spontaneously form a hierarchical array of FCDs, where each FCD acts as an individual gradient refractive index microlens. \linebreak[4]

Compared to microlens arrays reported in literature, our  ``compound eye'' SmA LC array possesses several advantages: 1) it is self-assembled by a curved LC-air interface in one step, which is simple and mass producible; the radius of curvature of the  ``compound eye'' can be tuned by the curvature of the interface and the shape of the pillars; 2) the lenses have a gradient size distribution and eccentricity from the pillar edge (high curvature) towards the regions between pillars (low curvature), thus, the number of lenses in each  ``compound eye'' and the spatial layout of lenses will be highly dependent on curvature of the interface; 3) it has variable DOF, allowing for 3D reconstruction of the composite images; 4) it is sensitive to light polarization; 5) it is stable at room temperature, but can be reconfigured when heated above the phase transition temperature; the optical properties (focal length, polarization) can be maintained over heating/cooling cycles.\linebreak[4]

\begin{figure}[!h]
\centering
\includegraphics[width=0.90\textwidth]{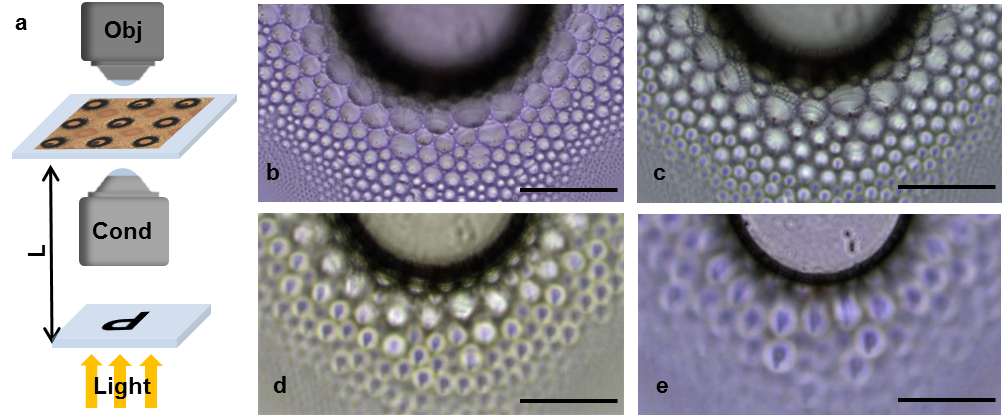}
\caption{Lensing effect of smectic FCDs in the flower pattern. a, Schematic illustration of the experimental set-up, indicating the light propagation direction (yellow arrows) and the relative positions of the light condenser (Cond), sample stage (where the top-down view of the flower pattern is shown), and objective (Obj). The letter P written on a microscope slide is placed at a distance of L$\approx$140 mm from the microscope stage. b, Flower pattern with the largest FCD in focus. c-e, The letter P imaged through the smectic microlenses at different focal planes. Shown are the focal planes where the smallest letters P are visible far from the pillar (c), where larger letters P are clearly visible (d), and where the largest letters are visible and many are distorted (e). Scale bars are 50 $\mu$m.}
\label{fig:stretchdef}
\end{figure}

Once we assembled the flower pattern (see details in the Experimental section, Supplementary Note 1 and Figure S1), we tested the lensing ability of the FCDs. To do this, we drew a letter  ``P'' (about 5 mm x 3mm) in black ink on a transparent microscope slide and placed it directly on top of the microscope lamp, roughly 140 mm away from the microscope stage (see schematic in Figure 1a). When the flower pattern is in focus, we do not see the letter P (Figure 1b). However, as we move the focus of the microscope objective above the plane of the FCDs, many small, inverted images of the letter P appear in correspondence with the position of each FCD: that is each FCD acts as a converging lens, resulting in an upside-down real image in agreement with literature \cite{kimmicrolens1, kimmicrolens2, land}. 
The image is formed at the focal plane of the microlenses. As the focus of the objective is moved farther above the microlenses, progressively bigger Ps appear in focus closer to the micropost (Figure 1c-e). The same effect can be observed when instead of imaging a letter we look at the focusing of light (supplementary Figure S2). \linebreak[4]

It is known that the focal length of FCDs depends on their size \cite{kimmicrolens1, kimmicrolens2}. In our system, the hierarchical arrangement of FCDs is dictated by the curved interface pinned at the edge of micropillars (visible in Figure 2a). In this self-assembled structure, most of the small FCDs are far away from the pillars, whereas progressively larger FCDs are found near the pinned edge. Small FCDs are also found close to the pillar edge because they improve the packing of the larger domains, as in Apollonius tiling \cite{meyer}. The range of FCD diameters is approximately 3-40 $\mu$m, which is remarkably similar to the ommatidia in insects \cite{land, barlow}. The total number of FCDs detected around each pillar amounts to a few hundred ommatidia, a number comparable to the 800 ommatidia in Drosophila \cite{strutt} and 1000 ommatidia in the horseshoe crab \cite{battelle}. Figure 2b shows the fraction of area occupied by FCDs in various concentric regions around the pillar (see details in Supplementary Note 2). The packing of FCDs is more efficient with medium-sized FCDs at a distance of about 60 $\mu$m from the center of the micropillar (radius 50 $\mu$m). In this region, the lenses also give the highest image quality. The large FCDs (> 20 $\mu$m) near the pillars are highly deformed and produce rather blurry or distorted images (Figure 1d). The bending of the smectic layers around the core results in highly deformed images, while the blurriness is due to the birefringence of the liquid crystal medium\cite{kimmicrolens1}. \linebreak[4]

\begin{figure}[!h]
\centering
\includegraphics[width=0.45\textwidth]{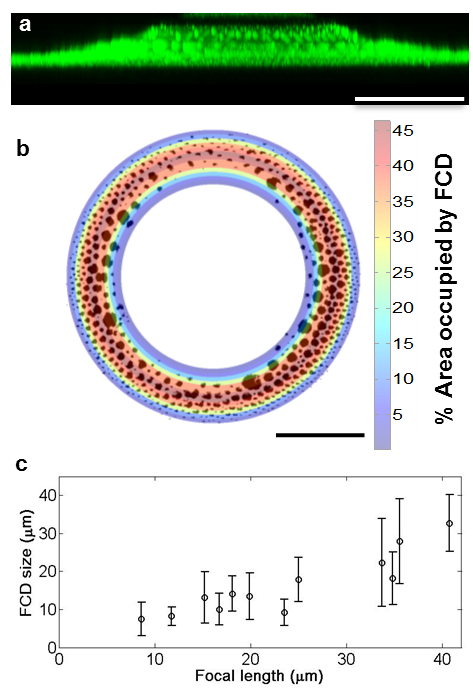}
\caption{Characterization of the distribution of FCD sizes and relative focal lengths in the flower pattern. a,  A stack of confocal images gives the profile of the interface of a flower pinned at the edge of a micropillar. b, The flower is divided into concentric regions around the pillar. In this diagram, the FCDs are represented as black spots. For each region, we extracted the total area occupied by FCDs, thus showing that a region exists where the FCDs are more efficiently packed. c, Relationship between the size of FCDs in a flower and the focal length. Scale bars are 50 $\mu$m.}
\label{fig:stretchdef}
\end{figure}

The relationship between the size of the FCDs and their focal length is plotted in Figure 2c. We note that the measured focal length of our FCDs is significantly larger than literature values \cite{kimmicrolens1, kimmicrolens2}  (10-40 $\mu$m vs. 1-3 $\mu$m), suggesting a possible role of the surface curvature on the focal length. Measurements of different assemblies of FCDs on various curved surfaces support this hypothesis (see Supplementary Note 3 and Figure S3).  \linebreak[4]

The absolute height H of a focused image is the sum of two lengths: the vertical position h of the FCD along the curved interface and its focal length f, which is proportional to its lateral size. Far from the pillar, where h is smallest, f is also smallest. As the curvature becomes steeper, both h and f (on average) increase. Hence, the separation $\Delta$H between the planes where the images are in focus for the smallest and the largest FCDs (see Figure S4) is maximized. However, if the light is shone above the pillars and collected below, $\Delta$H may become be smaller, as $\Delta$f and $\Delta$h would have a different sign: this opens the possibility of designing an array of microlenses on a curved interface whose focal planes all coincide. Such a system could be easily integrated with a flat array of light-guides or light collectors.   \linebreak[4]
One of the major attributes of LCs is their reconfigurability when heated above the phase transition temperature. Our  ``compound eye'' can be repeatedly generated and melted during cooling and heating cycles while maintaining the same lensing characteristics despite slight variations in FCD arrangement from cycle to cycle (see Videos V1 and V2). Such reconfigurability is distinctly superior to conventional lenses and any other artificial compound eye reported in literature \cite{lee, rogers}, whose focal lengths are not tunable or limited by the achievable mechanical compression. In addition, lenses that are damaged or distorted because of mechanical trauma can be reset here via a heating and cooling cycle.  \linebreak[4]

To test the possibility of superposing images using our flower lenses, we placed two different objects at different DOF from the microlenses: a letter  ``X'' (as in Figure 1a) and a photomask with dots (25 $\mu$m diameter) placed directly below the sample (see Figure 3a). Here, the pillars are fabricated on a coverslip (150-200 $\mu$m thick) in order to minimize the distance between the photomask and the sample. As shown in Figure 3b-c, while the letter  ``X'' appears on all lenses, the dots appear only in the lenses that are directly above, because each microlens can only sense a small part of the space, much like the apposition compound eye. In the apposition compound eye, each ommatidium is a lens that probes a small portion of space, conveys it to a waveguide and, finally, to a detector \cite{land}. The reproduction of this scheme has been the goal of many studies \cite{lee, rogers}. However, in a more primitive type of eye, the so-called schizochroal eye (present in ancient animals like the trilobites) \cite{fordyce, schon}, each ommatidium forms a whole image of a larger portion of the space, as do the FCDs in the flower pattern. Figure 3b shows that some have one dot in the center, some have a dot on the side, and others show two dots. \linebreak[4]

\begin{figure}[!h]
\centering
\includegraphics[width=0.45\textwidth]{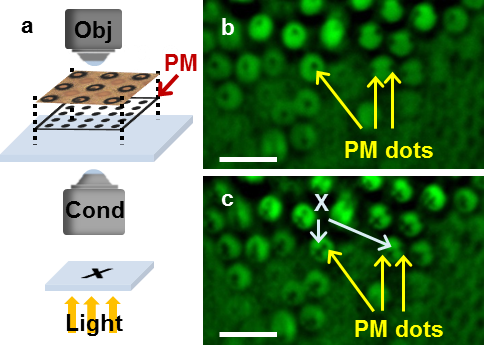}
\caption{Multiple objects in focus through the microlens array. a, Illustration of the optical set-up. The set-up is the same as in Figure 1a, but in addition a photomask (PM) of dots in a square array was inserted just beneath the sample, on top of the sample stage. The actual distance between the smectics and the photomask is therefore only the thickness of the coverslip that supports the smectic layers. b-c, Optical images showing (b) only the images of the photomask with dots (PM dots), some of which are highlighted by arrows and (c) the combined image of both the dots and the letter  ``X''. Scale bars are 50 $\mu$m.
}
\label{fig:stretchdef}
\end{figure}

Another important feature of natural compound eyes is the sensitivity to the direction of light polarization, which helps insects, for example, to distinguish petals\cite{fosterbio} and navigate their flight \cite{homberg}. The optical anisotropy of LCs make them perfect to achieve polarization sensitivity in our mimic compound eye. In fact, we show that the largest microlenses in our system, with the highest eccentricity, modify their lensing behavior with light polarization. We define the direction of the displacement of the FCD defect core with respect to the geometrical center of the FCD ellipse \cite{DeGennes, beller}, as the  ``eccentricity direction''. In the flower pattern, the eccentricity direction is radial from the center of the pillar (Figure 4a-b). We observe the lensing effect by projecting an object (a smiley face) in polarized light. As seen in Figure 4, vertically (North-South) polarized light blurs the images of the FCDs with vertical eccentricity, while images are clear with horizontally (East-West) polarized light. Vice versa, FCDs with horizontal eccentricity form images only with vertically polarized light. The polarization that matches the eccentricity also blurs the image. Video V3 clearly shows that as the polarizer rotates we can only see images projected from FCDs in certain positions, on top vs. bottom, and on the left vs. right with respect to the pillars. \linebreak[4]

\begin{figure}[!h]
\centering
\includegraphics[width=0.45\textwidth]{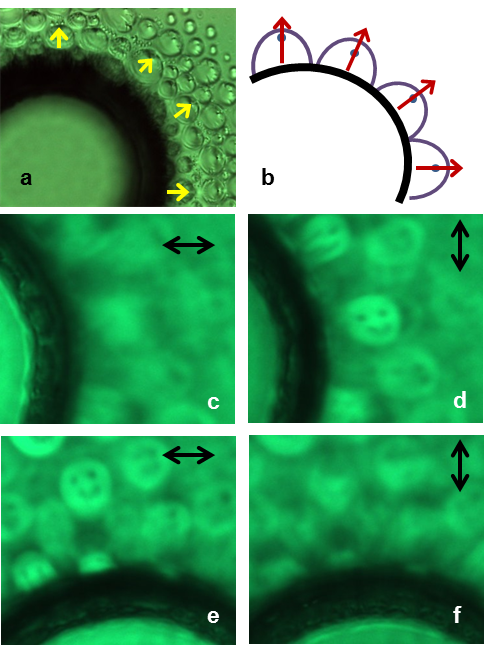}
\caption{Response of the microlenses to polarized light. In this case, the object is a smiley face. a, The bright field image of the flower and (b) the corresponding schematic illustrating the FCDs closest to the pillar and the position of their defect core, which determines the direction of the eccentricity, as indicated by the arrows. c-d, Polarized optical images showing the right side of the pillar and the flower pattern in bright field with two different polarizations. Here, the double-headed arrows indicate the direction of light polarization. The smiley faces appear only in d. e-f, Polarized optical images showing the top side of the pillar for two different polarizations. The smiley faces appear only in e. Scale bars are 30 $\mu$m. }
\label{fig:geom}
\end{figure} 

So far, the curvature of the compound eye layout is limited by the template to mold the lenses on. It is shown that hemispherical compound eye layout offers wide FOV without off-axis aberration \cite{rogers}.  In our system, by controlling the height and shape of the micropillars, the pinning strength of the SmA LC and the anchoring conditions, we can assemble FCDs not only in the regions of high negative curvature around the pillars, but also in positive curvature regions on top of the pillars and in saddle-points between the pillars  (Figure 5). Further, by tuning the spacing between pillars, we create a complete  ``garden'' of microlenses from a micropillar array (Figure 5a-b).  \linebreak[4]
 
\begin{figure}[!h]
\centering
\includegraphics[width=0.90\textwidth]{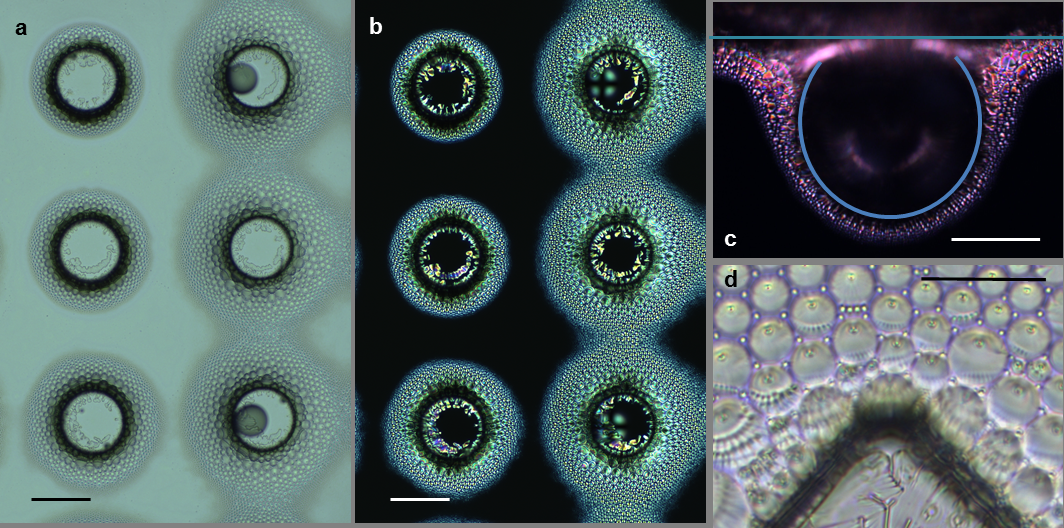}
\caption{ Tuning the assembly of FCDs. a-b, Bright field image (a) and polarized image (b) showing isolated pillars on the left (surface with negative curvature) and connected pillars on the right. The FCDs halfway between connected pillars are on a saddle point. c, A different assembly of FCDs between a pillar and a wall (whose position is approximately indicated by the blue line), with modulation of the size of the FCDs. d, Detail of the regular arrangement of FCDs near the corner of a square-base pillar. Note the large FCD at the vertex.  Scale bars are 100 $\mu$m in a, b and 50 $\mu$m in c, d. }
\label{fig:3chan}
\end{figure}

To conclude, we demonstrate a single-step approach to self-assemble  ``compound eye'' like smectic liquid crystal microlens arrays surrounding a single micropillar or micropillar arrays. These microlenses have wide FOV and DOF to construct 3D images.  They are highly reconfigurable and polarization sensitive. The melting and regeneration of the FCD flowers can be repeated many times without altering the optical characteristics. Therefore, our lenses are highly robust and self-healable, making them superior to existing artificial compound eye systems. Importantly, our approach is simple yet versatile. We can fine-tune the curvature of interface and consequently the layout of the compound eye by varying the surface anchoring of SmA LCs, the height, shape and arrangement of the pillars.  By integrating them into a more complex electronic circuitry, it is possible to create an optical device that will not only image, but sense and track the surroundings.  \linebreak[4]

\section{Experimental Section}

{\it Fabrication of  ``flowers'':} We first fabricated the SU-8 (MicroChem) micropillars with diameter of 100 $\mu$m, pitch of 300 $\mu$m, and height of 50 $\mu$m (see details in Supplementary Note 1 and Figure S1) using a photolithographic technique \cite{honglawan1, honglawan2, marcello} on a glass substrate to direct the assembly of SmA LCs, 4-cyano-4\'-octylbiphenyl (8CB, Kingston Chemistry). Then, we heated the substrate and the liquid crystals on a hot plate at 45$\deg$C into the isotropic phase. We then dropped a small amount of 8CB at the center of the pillar array and spread the liquid crystal evenly among the pillars using a glass coverslip. Thereafter, we slowly cooled the sample to room temperature. SU-8 imposes a degenerate planar anchoring and air imposes a perpendicular (homeotropic) anchoring with 8CB. As the temperature decreased and the LC passed from the isotropic to the smectic phase, a hierarchical assembly of FCDs gradually formed. 

{\it Optical Microscopy:} The measurement of the lensing effect from focal conic domains was performed with a Zeiss Axio Imager M1m optical microscope equipped with rotating polarizers. Images were acquired with a high-resolution camera (Zeiss AxioCam HRc). Objective magnification ranged from 50x to 100x. The condenser was adjusted for Koehler illumination. The focal length was measured by focusing first on a small group of focal conic domains and then adjusting the focus knob until the letters projected by that particular group of focal conic domains appeared sharp. The z-position of the microscope stage was monitored with an automatic controller. The difference between the two z-positions gave the focal length. For each image taken at a different z position we measured the size of all the FCD that gave well-focused images at that z position. The FCD size average was plotted against the focal length. Once the letters were in focus, small adjustment to the condenser lens allowed us to make the image appear more clearly or better illuminated, but did not detectably change the focal length. 
In some experiments, a green filter was used in order to observe the lensing behavior more clearly and avoid the problem of chromatic aberration.

{\it Confocal Microscopy:} Confocal imaging allowed for the 3D reconstruction of the flower texture. The system was prepared with LC molecules doped with 0.01$\%$ weight BTBP (N,N'-Bis(2,5-di-tert-butylphenyl)-3,4,9,10-perylenedicarboximide, Sigma Aldrich). The flower texture was visualized with a Leica TCS SP8 confocal microscope using an excitation laser line at 488 nm. The images were captured by LAS software and post-processed with FIJI software for quantitative analysis.

\section{Acknowledgements}

We thank Nader Engheta for many stimulating discussions, Alison Sweeney and Juliette McGregor for insights into the world of invertebrate eyes. We thank Andrea Stout and Jasmine Zhao for help with confocal microscopy. This work was supported by the National Science Fundation (NSF) Materials Science and Engineering Center (MRSEC) Grant to University of Pennsylvania, DMR-1120901. I. B. L. and N. D. B. are funded by GAANN $\#$P200A120246. This work was partially supported by a Simons Investigator grant from the Simons Foundation to R. D. K.

Authors Francesca Serra and Mohamed-Amine Gharbi contributed equally to this work.

\bibliographystyle{plain}

\clearpage

\beginsupplement 

\section*{Supplementary Material}

\subsection*{Supplementary Note 1: pillar fabrication}

Epoxy-based negative photoresists SU-8 (MicroChem Corp.) and KMPR (MicroChem Corp.) assure the random planar alignment of 8CB, which is necessary for the hybrid boundary conditions. Both materials are used to pattern the substrate. SU-8 and KMPR are fabricated on a glass substrate following standard lithographic procedures. First, clean glass slides (Fisherfinest, Fisher Scientific) are plasma etched for 60 seconds prior to lithography. The epoxy resin SU-8 or KMPR is spin-coated onto the glass substrate and exposed to 365 nm UV light (OAI Hybralign Series 200) through a photomask (CAD/Art Services, Inc). Thereafter, the sample is hard baked at 100$\deg$C for 6 minutes to crosslink the exposed regions and form. Subsequently, the glass substrate is placed in a developer solution (1-methoxy-2-propyl acetate, MicroChem Corp.) to dissolve the unexposed regions and leave the SU-8/KMPR pillar arrays. The surface details of the pillars are shown in Figure S1.

\subsection*{Supplementary Note 2: size distribution characterization}

A Matlab algorithm developed by us was used to measure the percentage of the total area occupied by FCDs at various radial distances from the center of the micropost. The brightness and contrast of a z-projection of the confocal stack were adjusted in FIJI, an open-source platform for biological image analysis and particle tracking, to obtain high contrast between the FCDs and the background while maintaining object size and shape. The algorithm first binarized the image and then fit circles to each object with area larger than a noise threshold. Then, equally spaced concentric annuli centered on the feature were generated. The number of pixels occupied by fit circles divided by the total number of pixels in each annulus was plotted in the heat map.

\subsection*{Supplementary Note 3: lensing effect from various interfaces}

We tested different assemblies of FCDs formed around various types of features, including conical pillars (50 $\mu$m in diameter on top, 25  $\mu$m in diameter on the bottom, 20  $\mu$m in height, 100  $\mu$m in pitch), spherical colloidal particles (10  $\mu$m in diameter), square pillars (100  $\mu$m sides, 50  $\mu$m in height, 300  $\mu$m in pitch), a spherical droplet on top of a wide cylindrical pillar (100  $\mu$m in diameter, 50  $\mu$m in height, 300  $\mu$m in pitch), and a flat substrate with hybrid boundary conditions (homeotropic at the air-smectic interface and degenerate planar at the smectic-substrate interface). In all cases, the focal length was linearly proportional to FCDs size (see Figure S3). However, the constant of proportionality between these two quantities was largest where the curvature gradient is the greatest (isolated flowers around large pillars far away from other pillars), supporting a possible curvature effect on the focal length. Curvature fields are controlled by the height of the micropillar and the thickness of the smectic film.

\clearpage

\subsection*{Supplementary figures}

\begin{figure}[!ht]
\centering
\includegraphics[width=0.90\textwidth]{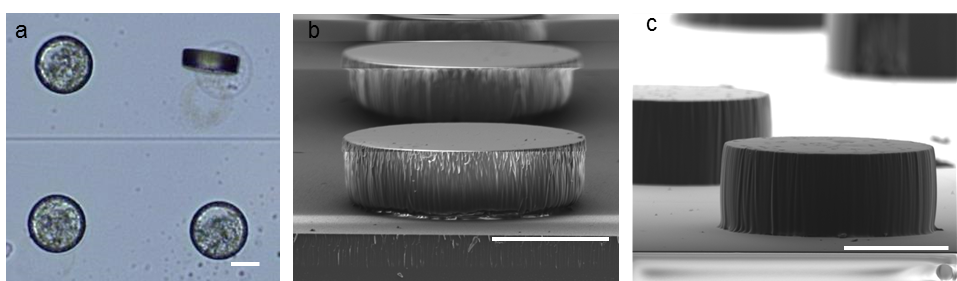}
\caption{Pillar roughness and shape characterization. a, Bright field optical images of the pillar. The side view under bright field microscopy is examined by gently scraping a pillar to remove it from the substrate and flip it onto its side. b-c, Scanning Electron Microscopy (SEM) images of (b) KMPR and (c) SU-8 pillars, revealing the roughness on the side of the pillars, prepared by photolithography. The substrate is tilted 45 degrees to capture these images. The scale bars are 50 $\mu$m.}
\label{fig:3chan}
\end{figure}

\begin{figure}[t!]
\centering
\includegraphics[width=0.90\textwidth]{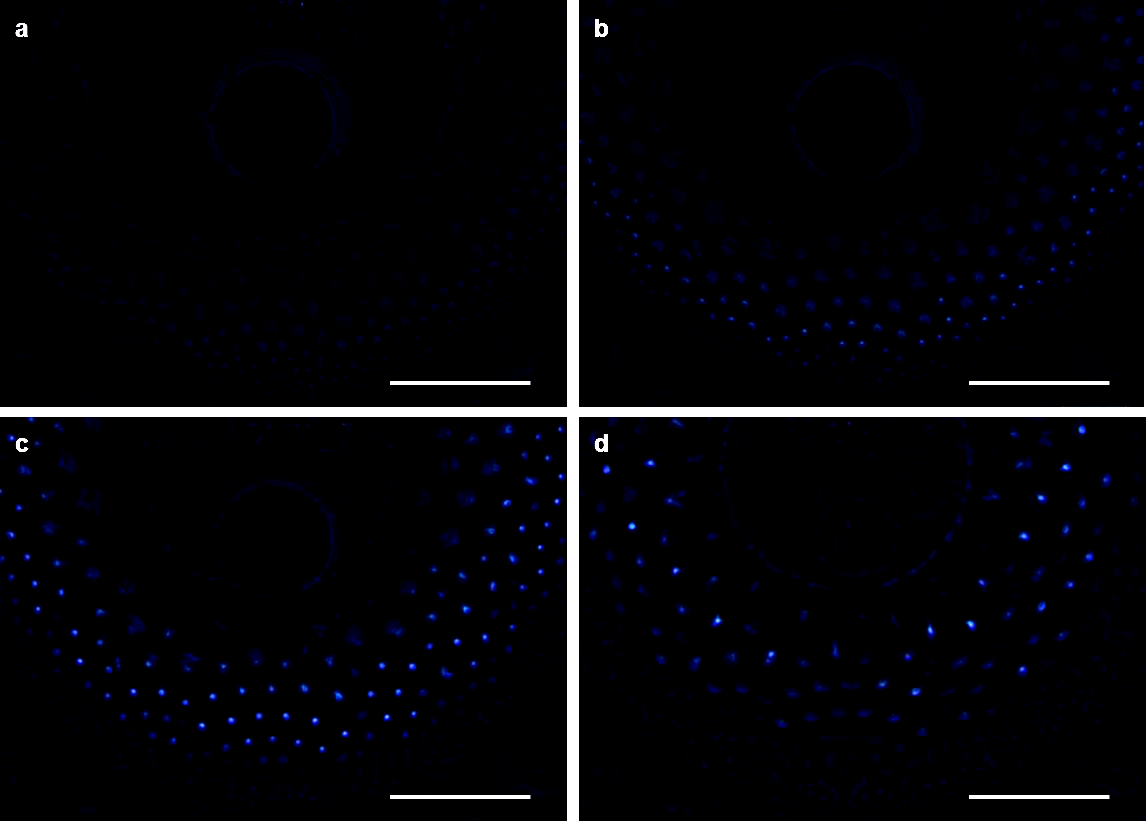}
\caption{Light focusing effect. a, The aperture diaphragm of the optical microscope was almost completely closed in order to lower the intensity of the incident light. As a result, the sample appears very dark. b-d, Although the sample is illuminated minimally, light is focused by different groups of microlenses as the plane of focus moves above the sample. The light focusing effect is additional proof that our system acts as a lens. The scale bars are 50 $\mu$m. }
\label{fig:3chan}
\end{figure}

\begin{figure}[t!]
\centering
\includegraphics[width=0.9\textwidth]{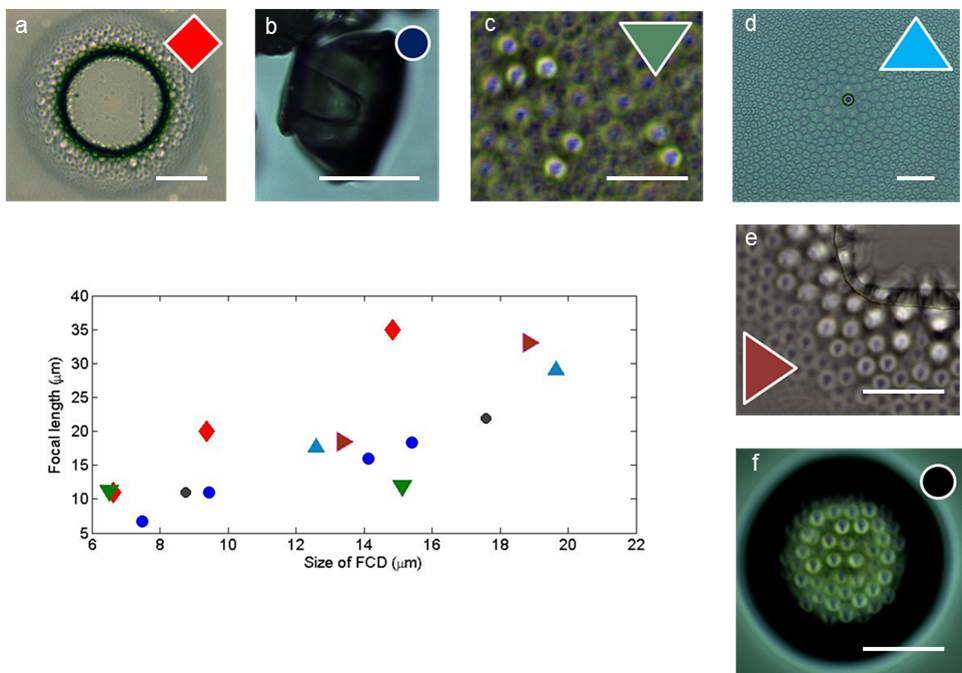}
\caption{ Focal lengths of different types of lenses. A number of other structures are used to sculpt the free surface of the smectic film and to promote the hierarchical assembly of FCD lenses. The lensing properties of FCDs are examined using the method specified in Supplementary Note 3 formed a, around an isolated circular pillar, b, around a tea-cup shaped pillar, c, on a flat substrate, d, around a single spherical colloid, e, around a square pillar, and f, in regions of positive curvature within a sessile drop. Even though the FCD distributions from a-f arise from a wide range of shapes and curvatures, their focal lengths vary with size. The scale bars are 50 $\mu$m. }
\label{fig:3chan}
\end{figure}

\begin{figure}[t!]
\centering
\includegraphics[width=0.55\textwidth]{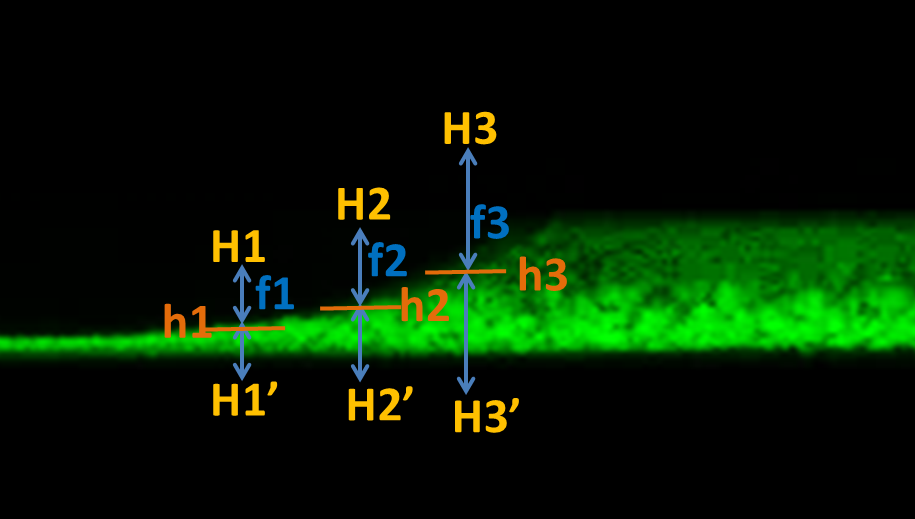}
\caption{Focal lengths. Due to the cylindrical symmetry of the system, FCDs at a given distance from the center have approximately the same sizes and eccentricities. The family of smallest FCDs is located in regions of smallest curvature at z-position $h_1$, with focal length $f_1$. The families of medium and large FCDs, at heights $h_2$ and $h_3$, have focal length $f_2$ and $f_3$ respectively, with $f_3 > f_2 >f_1$. The focal planes corresponding to these focal lengths ($H_1$, $H_2$, $H_3$) are maximally separated. If, instead, light is shone from above, the difference in height partly compensates for the difference in focal length, and the images would form at the planes $H_1'$, $H_2'$, $H_3'$, where $\Delta H'\ll\Delta H$.  }
\label{fig:3chan}
\end{figure}

\clearpage

\subsection*{Description of supplementary videos}

{\bf Supplementary Video V1.} Phase transition of reconfigurable lenses: melting. Melting of the smectic liquid crystals into the nematic phase (above 307 K) leads to the disappearance of FCDs. In this video, the microscope objective is focused on the focal plane of a group of microlenses (it is possible to distinguish some letters P) near the corner of a square pillar. 

{\bf Supplementary Video V2.} Phase transition of reconfigurable lenses: cooling. After the SmA phase is melted into the nematic phase (Supplementary Video 1) the sample is cooled back into the smectic phase without changing the focus of the microscope objective. The FCDs rearrange into the flower pattern and the lenses, as well as the letters P, are recovered. Hence, the lenses are temperature-tunable.

{\bf Supplementary Video V3.} Polarization effects of the FCD lenses. The flower is initially in focus (00:00). The focus is then shifted to see the images ("smilies") coming from the largest FCDs (00:01). The linear polarizer is initially in the vertical (north-south) position. The images are clearest on left and right. The linear polarizer from vertical to horizontal (images are clearest on top and bottom) and to vertical again, always causing a disappearance of the images that match the direction of the polarizer (00:03-00:05). FCD lenses are therefore polarization-sensitive.

\end{document}